\documentclass[journal]{IEEEtran}
\IEEEoverridecommandlockouts

\usepackage{titlesec}

\titlespacing*{\section}
  {0pt}{1.5ex plus 0.5ex minus 0.2ex}{0.8ex plus 0.2ex}

\usepackage{cite}
\usepackage[justification=centering,skip=4pt]{caption}
\usepackage{algpseudocode}
\usepackage{lettrine}
\usepackage{algorithm}
\usepackage{amsmath,amssymb,amsfonts}
\usepackage[justification=centering,skip=2pt,font=footnotesize]{caption}

\usepackage{mathtools}
\usepackage{graphicx}

\newtheorem{lemma}{Lemma}
\newtheorem{proof2}{Proof}

\newtheorem{proposition}{Proposition}

\newcommand{\C}{\mathbb{C}}
\usepackage{tabularx,booktabs}
\usepackage{textcomp}
\usepackage{xcolor}

\def\BibTeX{{\rm B\kern-.05em{\sc i\kern-.025em b}\kern-.08em
    T\kern-.1667em\lower.7ex\hbox{E}\kern-.125emX}}

\makeatletter

\def\tightdisplay{%
  \setlength\abovedisplayskip{2pt plus 1pt minus 1pt}%
  \setlength\belowdisplayskip{2pt plus 1pt minus 1pt}%
  \setlength\abovedisplayshortskip{0pt plus 1pt}%
  \setlength\belowdisplayshortskip{1pt plus 1pt minus 1pt}%
}
\g@addto@macro\normalsize{\tightdisplay}
\g@addto@macro\small{\tightdisplay}
\g@addto@macro\footnotesize{\tightdisplay}
\AtBeginDocument{\tightdisplay}
\def\section{\@startsection{section}{1}{\z@}%
  {0.8ex plus 0.4ex minus 0.2ex}{0.4ex plus 0.1ex}%
  {\normalfont\normalsize\centering\scshape}}
\def\subsection{\@startsection{subsection}{2}{\z@}%
  {0.6ex plus 0.3ex minus 0.2ex}{0.2ex plus 0.1ex}%
  {\normalfont\normalsize\itshape}}

\makeatother
\IEEEoverridecommandlockouts

\begin{document}

\title{Polyhedral RIS Architecture with Quantum-Assisted Facet Selection for Uniform Coverage}

\author{Noha~Hassan,~\IEEEmembership{SMIEEE,}
        and~Halim~Yanikomeroglu,~\IEEEmembership{FIEEE}%
\thanks{This work has been submitted to the IEEE WCL for possible publication. Copyright may be transferred without notice, after which this version may no longer be accessible.}
\thanks{N. Hassan and H. Yanikomeroglu are with the Department of Systems and Computer Engineering,
Carleton University, Ottawa, ON, Canada. Computer Engineering, Dalhousie
University, NS, Canada.
(e-mails: nohahassan@sce.carleton.ca; noha.hassan@dalca; halim@sce.carleton.ca).}
}
\maketitle

\markboth{IEEE Wireless Communications Letters}%
{Hassan \MakeLowercase{\textit{et al.}}: }
\pagestyle{headings}

\maketitle

\begin{abstract}
Reconfigurable intelligent surfaces (RISs) direct coverage toward moving user equipment (UE), but practical deployments involve discrete phase control, uncertain facet positions, and UE mobility, which require the problem to be re-solved continually. We consider these challenges for the chamfered hexagonal polyhedral RIS, which provides nearly uniform coverage. With both incidence and departure angles, we bound in closed form the gain loss of its best facet relative to the best possible reflecting facet, include the known $b$-bit quantization loss, and extend the bound to joint random mounting errors of all facets. For a precoded multi-user downlink, each facet-selection subproblem of the max-min rate design is shown to be an exact quadratic unconstrained binary optimization (QUBO) problem , suited to repeated time-constrained solving. The deployment of the base station in relation to the RIS allows the polyhedral surface to bring down the percentage of dead zones in vertical planar panels from 77\% to 0.3\% and increase the minimum rate by 34\% relative to the STAR panel. The simulated annealing algorithm finds the optimum solution in 95\% of the cases. The quantum approximation algorithm fails to do so.
\end{abstract}

\vspace{-0.3em}
\section{Introduction}
\lettrine{S}{ixth generation} (6G) wireless networks have been more focused on utilizing reconfigurable intelligent surfaces (RIS) for controlling wireless propagation and enhancing network coverage. Using double-sided RIS panels increases the coverage area compared to single-sided RIS. Nevertheless, a single-sided panel serves only one half-space \cite{gao2025dual}. Although double-sided and simultaneously transmitting and reflecting (STAR) panels serve both half-spaces \cite{xu2022simultaneously,mu2022star}, their gain rolls off toward grazing angles, so users close to the panel plane see a deep loss. This drawback is  relevant to both high-rise systems and the integrated sensing and communication (ISAC) context, where it becomes common practice for the user equipment (UE) to be present at high elevation angles that are unattainable by a simple fixed-panel normal.
The adoption of the polyhedral surface geometry, which is achieved through a chamfered hexagonal prism with controllable facets, resolves this issue by providing many reflecting normals as opposed to a single fixed-panel one, yet at the cost of the introduction of a facet activation design variable unique to the polyhedral surfaces. We formulate this discrete facet selection problem as a quadratic unconstrained binary optimization (QUBO) problem and incorporate it into the quantum polyhedral RIS optimizer (QPRO), which uses a hybrid quantum-classical strategy in which the phase and spacing parameters are refined classically.

The main contributions of this paper are the following: (1) We derive a worst-case directional gain guarantee with both incidence and departure angles that includes the known quantization loss \cite{wu2020discrete}. (2) We extend the deterministic worst-case directional gain guarantee of the underlying polyhedral geometry to a probabilistic bound under random facet-mounting perturbation  of all facets jointly, characterizing how gracefully the guarantee degrades under realistic installation tolerances. (3) We show that with precoding and coupling, each facet selection subproblem of the max-min design is an exact QUBO, which is suited to repeated, time-constrained re-solving as UEs move between coherence intervals, and validate it using Monte Carlo simulations via a classical simulated annealing approach  against the exhaustive optimum. (4) Noiseless QAOA and real quantum computer are compared to classical algorithms for feasibility and optimality gaps.
\vspace{-0.3em}
\section{Related Work}
\vspace{-0.1em}
\label{sec1}
The RIS technology has matured from a phase-only supplementary approach to a full-fledged field of research that includes various hardware platforms \cite{basar2024reconfigurable}, advanced meta-surfaces, and RIS-based localization \cite{umer2025reconfigurable}. Studying the issue of impedance-based restrictions on symmetric reflection-transmission behavior \cite{xu2022simultaneously}, sum-rate-based beamforming, power control, and scheduling \cite{gao2025dual}, backscattering-based switching where variable switching modes may be used to enhance the modulation rate in the moderate-high signal-to-noise ratio (SNR) regime \cite{huang2025riding}, cooperative relaying with multiple panels for vehicle-to-vehicle (V2V) communications \cite{ji2026dual}, and increasing the multi-UE capacity through extra spatial degrees of freedom due to the two-sided panel structure \cite{peng2025capacity}, followed by cross-layer resource management \cite{zhou2023queueing}.
Nonetheless, all these efforts remain confined within two dimensions; using several panels to cover all directions only adds deployment cost, complexity, and mutual interference \cite{ji2026dual,peng2025capacity}, without resolving the underlying geometric limitation.
The iterations of alternating optimization (AO) update the beamformers based on the phase shifts of the RIS using semidefinite relaxation, successive convex approximation, or fractional programming \cite{xiu2021reconfigurable}. Greedy update rule heuristic is used with AO for large antenna arrays without guaranteeing optimality \cite{11404429}, and dual simultaneously transmitting and reflecting (D-STAR) methods use AO to optimize the transmission and reflection coefficients of the two-sided panels while showing improved spectral efficiency but solving sum-rate optimization instead of worst-case omnidirectional coverage optimization problem solved in this paper \cite{10431763}. None of these algorithms incorporate the facet selection binary variable. The binary coupled constrained combinatorial optimization problem in this paper is unique to multi-facet polyhedron surfaces.
The physical hardware for the RIS can only process a finite set of phase states, for which closed-form or linear-time discrete beamforming algorithms exist \cite{ren2022linear,pekcan2024achieving}, and the worst-case loss of quantization levels is known \cite{wu2020discrete}. This motivates facet activation as a QUBO in our formulation rather than continuous relaxation rounding. There were also initial studies that examined the formulation of optimization problems associated with 6G baseband and resource allocation to quantum annealers and the practicality of doing so \cite{kim2021heuristic}. Recent research uses hybrid quantum methods to solve general optimization problems in network design and the broader intersection between quantum computing and wireless communication/networking \cite{zhao2024quantum}. However, this literature has not attempted to optimize a discrete facet selection variable, as it has not even considered the existence of a multifaceted surface.
\vspace{-0.2em}
\section{System Model and Polyhedral Architecture}
\label{sec2}
\begin{figure}[ht]
\centering
\includegraphics[width=2.2in, height=1.1in]{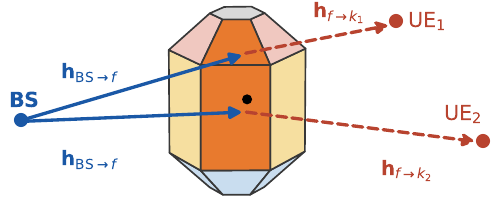}
\caption{System model of the chamfered polyhedral RIS.}
\label{poly}
\end{figure}

The surface of the RIS consists of $F$ facets of a plane geometry, where the facet $f$ has $N_f$ elements placed at spacings $\mathbf{d}_f=[d_{f,1},\ldots,d_{f,N_f}]$.
An orientation specified by a normal $\mathbf{n}_f$ corresponds to the elevation and azimuth angles $(\theta_f,\varphi_f)$, respectively. The independent phase vector is defined as $\boldsymbol{\phi}_f=[\phi_{f,1},\ldots,\phi_{f,N_f}]$.
The geometry considered here is that of a chamfered hexagonal prism having $F = 20$ facets distributed into three elevation bands (six vertical facets, six upper-beveled, and six lower-beveled facets, along with top and bottom caps). The $60^\circ$ bevel tilt (normals at $\theta_f=60^\circ$ and $\theta_f=120^\circ$) is chosen so that each bevel ring's $\pm30^\circ$ half-power span exactly closes the elevation gap between the vertical and cap facets, rather than leaving a dead zone near the horizon as a shallower or steeper chamfer would as shown in Fig. \ref{poly}.
Each facet pair in a particular band is separated by $60^\circ$ azimuth angle so that every direction lies within $\delta_c\approx33.7^\circ$ of its best aligned facet normal for nominal unperturbed mounting, where $\delta_c$ is the covering radius of the normal set.
In reality, each facet is mounted, which makes the actual normal $\mathbf{n}_f'=\mathbf{n}_f + \boldsymbol{\delta}_f$ of each facet deviate from its nominal orientation $\mathbf{n}_f$ above due to a small random mounting orientation error $\boldsymbol{\delta}_f$. The effect of such perturbation is analyzed in Section~\ref{sec3}. The UE locations are also slowly varying relative to a coherence time $T_c$ during which the phase and spacing parameters are considered fixed while the facet activation can be re-solved. The base station (BS) has $M$ antennas and illuminates the RIS from above, and its direct links to the $K$ UEs are obstructed.
\vspace{-0.3em}
\subsection{Channel Model}
The polyhedral RIS behaves as a single linear system. It takes the symbols that the BS transmits and maps them to what each UE receives. This mapping depends on which facets are switched on. For a given UE $k$, this mapping is expressed as
\begin{equation}
\label{eq:systemvec}
\mathbf{h}_k^H(\boldsymbol{\alpha}) = \mathbf{h}_{d,k}^H+\textstyle\sum_{f=1}^F\alpha_f\,\mathbf{h}_{f\to k}^{H}\,\Theta_f\,\mathbf{G}_{\mathrm{BS}\to f}^{\mathrm{eff}},
\end{equation}
where $\boldsymbol{\alpha}$ is the facet activation vector $\boldsymbol{\alpha}=[\alpha_1,\ldots,\alpha_F]^{T}\in\{0,1\}^F$, $\mathbf{h}_{d,k}\in\C^{M}$ is the obstructed direct link, $\mathbf{G}_{\mathrm{BS}\to f}^{\mathrm{eff}}\in\C^{N_f\times M}$ and $\mathbf{h}_{f\to k}\in\C^{N_f}$ are the BS-to-facet and facet-to-UE links, where inactive facets do not re-radiate.

In contrast to the passive reflector where the phase control is the only factor present, each element has its own local angle of incidence that determines the roll-off in the response function, so $\Theta_f$ represents both gain and phase, and is given by
\begin{equation}
\label{eq:elementgain}
\Theta_f = \mathrm{diag}\!\big(\cos^{p}\!\psi_{f,1}\,e^{j\phi_{f,1}},\ldots,\cos^{p}\!\psi_{f,N_f}\,e^{j\phi_{f,N_f}}\big),
\end{equation}
where $\psi_{f,n}$ is element $n$'s local misalignment to the incident wave, $p\geq1$ is the roll-off sharpness, and $\phi_{f,n}$ is the independently controlled phase. The same pattern applies on departure, so facet $f$ reaches UE $k$ with the pattern product $\rho_{f,k}=\cos^p\psi^{\rm in}_f\cos^p\psi^{\rm out}_{f,k}$ of its incidence and departure angles with zero beyond $90^\circ$. A facet that the BS does not illuminate does not contribute.

Facets sit at an angle to one another, and each scatters energy toward neighbors that re-enter through the same coupling path. The effective channel satisfies the fixed point relation as $\mathbf{G}_{\mathrm{BS}\to f}^{\mathrm{eff}} = \mathbf{G}_{\mathrm{BS}\to f} + \sum_{f'\neq f}\alpha_{f'}\mathbf{C}_{ff'}\Theta_{f'}\mathbf{G}_{\mathrm{BS}\to f'}^{\mathrm{eff}}$, considering that only active facets re-radiate.
It is also expressed in block form over all $F$ facets as
\begin{equation}
\label{eq:exactcoupling}
\mathbf{G}^{\mathrm{eff}} = \big(\mathbf{I} - \mathbf{C}\mathbf{D}_{\alpha}\boldsymbol{\Theta}\big)^{-1}\mathbf{G}_{\mathrm{BS}},
\qquad
\mathbf{C} = \begin{bmatrix} \mathbf{0} & \mathbf{C}_{12} & \cdots \\ \mathbf{C}_{21} & \mathbf{0} & \cdots \\ \vdots & & \ddots \end{bmatrix},
\end{equation}
where $\mathbf{D}_\alpha=\mathrm{blkdiag}(\alpha_f\mathbf{I}_{N_f})$ and $\boldsymbol{\Theta}=\mathrm{blkdiag}(\Theta_f)$. As $\|\mathbf{C}_{ff'}\|\ll1$, the spectral radius of $\mathbf{C}\mathbf{D}_\alpha\boldsymbol{\Theta}$ is below one, the Neumann series converges quickly, and only the first-order term is retained as
\begin{equation}
\label{eq:effchannel}
\mathbf{G}_{\mathrm{BS}\to f}^{\mathrm{eff}} \approx \mathbf{G}_{\mathrm{BS}\to f}+\textstyle\sum_{f'\neq f}\alpha_{f'}\,\mathbf{C}_{ff'}\Theta_{f'}\mathbf{G}_{\mathrm{BS}\to f'},
\end{equation}
\begin{equation}
[\mathbf{C}_{ff'}]_{ij} = \kappa_c\,\mathrm{sinc}(k_0 d_{ij})\,e^{-jk_0 d_{ij}}.
\end{equation}
Here each facet receives cross-facet terms from the active facets with coupling strength $\kappa_c$. Since $|[\Theta_f]_{nn}|\le1$, $\|\mathbf{C}\|_2<1$, this guarantees convergence for every activation and phase configuration. Here, $d_{ij}=\|\mathbf{r}_{f,i}-\mathbf{r}_{f',j}\|$ and $k_0$ is the free-space wavenumber.

Each entry of the direct link $\mathbf{h}_{\mathrm{BS}\to f}$ is Rician and is represented as
\begin{equation}
\label{eq:rician}
h_{\mathrm{BS},f,n} = \sqrt{\tfrac{\kappa}{\kappa+1}}\,h_{\mathrm{BS},f,n}^{\mathrm{LoS}} + \sqrt{\tfrac{1}{\kappa+1}}\,h_{\mathrm{BS},f,n}^{\mathrm{NLoS}},
\end{equation}
where the Rician factor is $\kappa$, $h_{\mathrm{BS},f,n}^{\mathrm{NLoS}}\sim\mathcal{CN}(0,1)$, and the LoS term is
$h_{\mathrm{BS},f,n}^{\mathrm{LoS}} = \frac{\lambda}{4\pi r_{f,n}} e^{-j\frac{2\pi}{\lambda}r_{f,n}} e^{j\frac{2\pi}{\lambda}\mathbf{k}(\theta_f,\varphi_f)\cdot\mathbf{r}_{f,n}}$. This carries the free-space path loss and the incidence-dependent spatial phase set by the propagation vector $\mathbf{k}(\theta_f,\varphi_f)$ and element location $\mathbf{r}_{f,n}$ with $r_{f,n}=\|\mathbf{r}_{f,n}-\mathbf{r}_{\mathrm{BS}}\|$. The same model applies to each BS antenna and to the facet-to-UE links.
\vspace{-0.1cm}
\subsection{Signal Model}
The BS transmits $\sum_{i}\mathbf{w}_ix_i$ with linear precoders $\mathbf{w}_i\in\C^M$ and $\sum_i\|\mathbf{w}_i\|^2\le P$. From \eqref{eq:systemvec}, UE $k$ receives $y_k = \mathbf{h}_k^H(\boldsymbol{\alpha})\sum_{i=1}^{K}\mathbf{w}_ix_i + n_k$, where $n_k\sim\mathcal{CN}(0,\sigma_k^2)$ is the additive noise, so the inter-UE interference reaches UE $k$ through its own channel and carries the other UEs' precoded streams.
The signal-to-interference-and-noise ratio (SINR) is defined as
$\mathrm{SINR}_k = \frac{|\mathbf{h}_k^H(\boldsymbol{\alpha})\mathbf{w}_k|^2}{\sum_{i\neq k}|\mathbf{h}_k^H(\boldsymbol{\alpha})\mathbf{w}_i|^2 + \sigma_k^2}$.
The data rate follows from SINR and is expressed as
\begin{equation}
R_k\big(\{\boldsymbol{\phi}_f,\mathbf{d}_f,\alpha_f\}_{f=1}^F,\mathbf{W}\big) = \log_2(1+\mathrm{SINR}_k).
\end{equation}

This rate depends on how well each facet is aligned to both the BS and the UE it serves. The element pattern terms cause $\rho_{f,k}$, and therefore $R_k$, to drop as alignment worsens.
\section{Problem Formulation and Algorithm Description}
\label{sec3}

\begin{proposition}
\label{prop:quant_loss}
With $\Omega_k$, which is the angle between the BS and UE directions, (i) any facet has $\rho_{f,k}\le\cos^{2p}(\Omega_k/2)$, with equality only for the bisecting normal; (ii) if $\Omega_k/2+\delta_c<90^\circ$, $\max_f\rho_{f,k}\ge\cos^{2p}(\Omega_k/2+\delta_c)$, a maximum loss is calculated as
\begin{equation}
\label{eq:quant_gain_bound}
L(\Omega_k,\delta_c)=40p\log_{10}\frac{\cos(\Omega_k/2)}{\cos(\Omega_k/2+\delta_c)}~\mathrm{dB}.
\end{equation}
\end{proposition}
\begin{proof2}
With $a,c$ as the angles of the normal to the BS and UE directions, then $a+c\ge\Omega_k$ and $\cos a\cos c\le\cos^2\frac{a+c}{2}\le\cos^2\frac{\Omega_k}{2}$, with equality only for $a=c=\Omega_k/2$. Hence, a facet within $\delta_c$ of the bisector has $a,c\le\Omega_k/2+\delta_c$.
\end{proof2}
Considering $b$-bit phases in \eqref{eq:phase}, the known bound $|g_{f,k}^{\mathrm{quant}}| \geq \cos(\pi/2^b)|g_{f,k}^{\mathrm{opt}}|$ \cite{wu2020discrete} adds $\Delta_b \;=\; -20\log_{10}\!\left(\cos\frac{\pi}{2^b}\right)~\mathrm{dB}$ to \eqref{eq:quant_gain_bound}, which vanishes as $b\to\infty$.
\begin{proposition}
\label{prop:robust}
Suppose that the realized normal of each facet has $\mathbf{n}_f' = \mathbf{n}_f + \boldsymbol{\delta}_f$, where $\boldsymbol{\delta}_f$ is an isotropic random perturbation in the mounting orientation with $\|\boldsymbol{\delta}_f\|$, its angular size, having a distribution with standard deviation $\sigma_\delta$ (in degrees), independently for each facet. Then, for any $\epsilon\in(0,1)$, the margin \eqref{eq:quant_gain_bound} with $\delta_c$ replaced by $\delta_c+\delta_\epsilon$ holds for all directions simultaneously with probability at least $1-\epsilon$,
where $\delta_\epsilon$ denotes the $(1-\epsilon)^{1/F}$ quantile of $\|\boldsymbol{\delta}_f\|$.
\end{proposition}
\begin{proof2}
From the triangle inequality on the sphere, a direction within $\delta_c$ of a nominal normal is within $\delta_c+\max_f\|\boldsymbol{\delta}_f\|$ of the realized one, for all directions simultaneously, and by independence $\Pr[\max_f\|\boldsymbol{\delta}_f\|\le\delta_\epsilon]=\Pr[\|\boldsymbol{\delta}_f\|\le\delta_\epsilon]^F=1-\epsilon$.
\end{proof2}
\subsection{Joint Optimization Problem}
Phase, spacing, precoding, and activation are optimized jointly for max-min UE rate according to the following objective function:
\begin{subequations}
\begin{align}
& \max_{\{\boldsymbol{\phi}_f,\mathbf{d}_f,\alpha_f\}_{f=1}^F,\mathbf{W}}\;
  \min_k\; R_k \label{eq:obj}\\
& \text{s.t.}\;\;\phi_{f,n}\in\{m\Delta\phi\}_{m=0}^{2^b-1},\;
  \Delta\phi=\tfrac{2\pi}{2^b},\;\forall f,n, \label{eq:phase}\\
& \hphantom{\text{s.t.}\;\;}\mathbf{d}_f\in[d_{\min},d_{\max}]^{N_f},\;\forall f, \label{eq:spacing}\\
& \hphantom{\text{s.t.}\;\;}\textstyle\sum_{n=1}^{N_f} d_{f,n}\le D_{\mathrm{total}}^{(f)},\;\forall f, \label{eq:aperture}\\
& \hphantom{\text{s.t.}\;\;}\textstyle\sum_{k=1}^K\|\mathbf{w}_k\|^2\le P, \label{eq:power}\\
& \hphantom{\text{s.t.}\;\;}\mathrm{SINR}_k\ge\gamma_{\min},\;\forall k, \label{eq:sinr}\\
& \hphantom{\text{s.t.}\;\;}\alpha_f\in\{0,1\},\;\forall f, \label{eq:binary}\\
& \hphantom{\text{s.t.}\;\;}\textstyle\sum_{f=1}^F\alpha_f\le F_{\max}, \label{eq:facetcount}\\
& \hphantom{\text{s.t.}\;\;}\alpha_f\alpha_{f'}\|\mathbf{C}_{ff'}\|_{\max}\le C_{\max},\;\forall f\ne f'. \label{eq:coupling}
\end{align}
\end{subequations}
Constraint \eqref{eq:phase} {restricts every element's phase shift to one of $2^b$ discrete levels, where $b$ is the number of phase control bits per element and \eqref{eq:spacing} keep every element's phase shift and spacing within a certain range, respectively. Constraint \eqref{eq:aperture} ensures that the elements on a facet do not exceed the facet's total aperture budget. Constraint \eqref{eq:power} limits the transmit power. Constraint \eqref{eq:sinr} guarantees a minimum quality of service for every UE  and holds when the optimal value of \eqref{eq:obj} is at least $\log_2(1+\gamma_{\min})$. Constraints \eqref{eq:binary} and \eqref{eq:facetcount} together decide which facets can be switched on and cap how many are active simultaneously. Finally, constraint \eqref{eq:coupling} limits the allowed electromagnetic coupling magnitude between any two active facets, forbidding a set $\mathcal{E}$ of pairs.}
This problem is non-convex, since the phase-dependent channel term inside $R_k$, the binary activation variables, and the distance-dependent coupling constraint all interact with one another.
\subsection{Facet Selection Algorithm Description}
In the previous joint problem, phase and spacing variables are continuous and can be refined with ordinary numerical methods, while activation is discrete and combinatorial.
This problem is solved using two stages: a classical stage that jointly refines variables to convergence as in stage 1 of Algorithm~\ref{alg1} and a separate hardware-verification stage that solves only the discrete step while freezing phase and spacing as in stage 2. Because UEs move between coherence intervals, the discrete step is intended to be re-solved once per interval. Stage 1 matches each facet with its UE (starting with the weakest UE) and co-phases it, employs  regularized zero-forcing (RZF) precoding, and conducts a search on spacing. Each of these updates is considered based on \eqref{eq:exactcoupling}, provided that the minimum rate has increased.
\begin{algorithm}[t]
\scriptsize
\algrenewcommand\algorithmicindent{0.8em}
\caption{Polyhedral Facet Selection Algorithm}
\label{alg1}
\begin{algorithmic}[1]
\State \textbf{Input:} $\{\mathbf p_k\},\{\mathbf h_{\mathrm{BS}\to f}\},\{\mathbf h_{f\to k}\}$;
\textbf{Init:} feasible $\boldsymbol\phi_f,\mathbf d_f$, $\alpha_f\!\leftarrow\!1$
\Repeat
\State $\mathbf{W}\leftarrow$ RZF; $\gamma\leftarrow\min_k\mathrm{SINR}_k$; $\boldsymbol{\mu}\leftarrow\mathbf{1}/K$ \Comment{Stage 1}
\For{$t=1,\ldots,T$ (Dinkelbach on $\gamma$)}
\State $\{\alpha_f\}\leftarrow\textsc{SolveQUBO}(H_{\mathrm{QUBO}})$ (SA/QAOA);
keep $\boldsymbol{\alpha}$ if $\min_k\mathrm{SINR}_k$ rises; update $\boldsymbol{\mu},\gamma$
\EndFor
\State Refine $\{\boldsymbol\phi_f,\mathbf d_f\}$ for $J$ with $\alpha$ fixed; project $\phi_{f,n}$ to levels in \eqref{eq:phase} \Comment{Stage 2}
\Until{convergence}
\State \textbf{Return:} $\{\boldsymbol\phi_f^*,\mathbf d_f^*,\alpha_f^*\}$; raw hardware statistics if applicable
\end{algorithmic}
\end{algorithm}
\begin{lemma}
\label{lem:quad}
For fixed $\Theta_f,\mathbf{d}_f,\mathbf{W}$, $\mathbf{h}_k^H\mathbf{w}_i=\mathbf{h}_{d,k}^H\mathbf{w}_i+\sum_f\alpha_f\mathbf{h}_{f\to k}^H\Theta_f\mathbf{G}_{\mathrm{BS}\to f}\mathbf{w}_i$ is linear in $\boldsymbol{\alpha}$, hence $g_k=|\mathbf{h}_k^H\mathbf{w}_k|^2-\gamma(\sum_{i\neq k}|\mathbf{h}_k^H\mathbf{w}_i|^2+\sigma_k^2)$ is quadratic in $\boldsymbol{\alpha}$, and $\gamma$ is achievable if and only if $\max_{\boldsymbol{\alpha}}\min_kg_k\ge0$.
\end{lemma}
A Dinkelbach update $\gamma\leftarrow\min_k\mathrm{SINR}_k$ and UE weights $\mu_k\leftarrow\mu_ke^{-\eta g_k/\max_i|g_i|}$ handle the max-min structure. With $\sum_k\mu_kg_k=q_0+\mathbf{q}^T\boldsymbol{\alpha}+\boldsymbol{\alpha}^T\mathbf{Q}\boldsymbol{\alpha}$ ($\mathbf{Q}$ upper triangular), the discrete step minimizes the following:
\begin{equation}
\label{eq6}
\resizebox{0.95\linewidth}{!}{$\displaystyle
H_{QUBO} = -\mathbf{q}^T\boldsymbol{\alpha}-\boldsymbol{\alpha}^T\mathbf{Q}\boldsymbol{\alpha}+\lambda\Big[\big(\textstyle\sum_f\alpha_f+\mathbf{c}^T\mathbf{s}-F_{\max}\big)^2+\textstyle\sum_{(f,f')\in\mathcal{E}}\alpha_f\alpha_{f'}\Big]
,$}
\end{equation}
where slack bits $\mathbf{s}$ with weights $\mathbf{c}=[1,2,\ldots,2^{S-2},F_{\max}-2^{S-1}+1]$, $S=\lceil\log_2(F_{\max}+1)\rceil$, let $\mathbf{c}^T\mathbf{s}$ take every integer in $[0,F_{\max}]$, so fewer active facets pay no penalty.

\begin{proposition}
\label{prop:qubo}
If $\lambda>\|\mathbf{q}\|_1+\sum_{f<f'}|Q_{ff'}|$, every minimizer of \eqref{eq6} satisfies \eqref{eq:facetcount}, \eqref{eq:coupling} and maximizes $\sum_k\mu_kg_k$ over them.
\end{proposition}
\begin{proof2}
The range of $J(\boldsymbol{\alpha})=\mathbf{q}^T\boldsymbol{\alpha}+\boldsymbol{\alpha}^T\mathbf{Q}\boldsymbol{\alpha}$ is at most $B=\|\mathbf{q}\|_1+\sum_{f<f'}|Q_{ff'}|$. A feasible $\boldsymbol{\alpha}$ has a slack with zero penalty, so $\min_{\mathbf{s}}H_{QUBO}=-J(\boldsymbol{\alpha})$. Otherwise, the bracket in \eqref{eq6} is a positive integer and $H_{QUBO}\ge-J(\boldsymbol{\alpha})+\lambda>-J(\boldsymbol{\alpha}^\star)$ for any feasible $\boldsymbol{\alpha}^\star$.
\end{proof2}
For $F_{\max}=8$, \eqref{eq6} has 24 variables. No QUBO is asserted to be equivalent to the max-min optimization problem, whereby each solution is evaluated based on the true minimum SINR and is benchmarked against the exhaustive optimal solution. Since an update will be accepted only if the true minimum SINR increases, and since the feasible set is finite, Algorithm~\ref{alg1} will terminate after a finite number of iterations. Each QUBO has $F + S$ variables, while one sweep through simulated annealing takes $\mathcal{O}((F + S)^2)$ time.
\vspace{-0.3em}
\section{Experimental Results}
\label{sec4}
\begin{figure}[ht]
\centering
\includegraphics[width=3.1in, height = 1.0in]{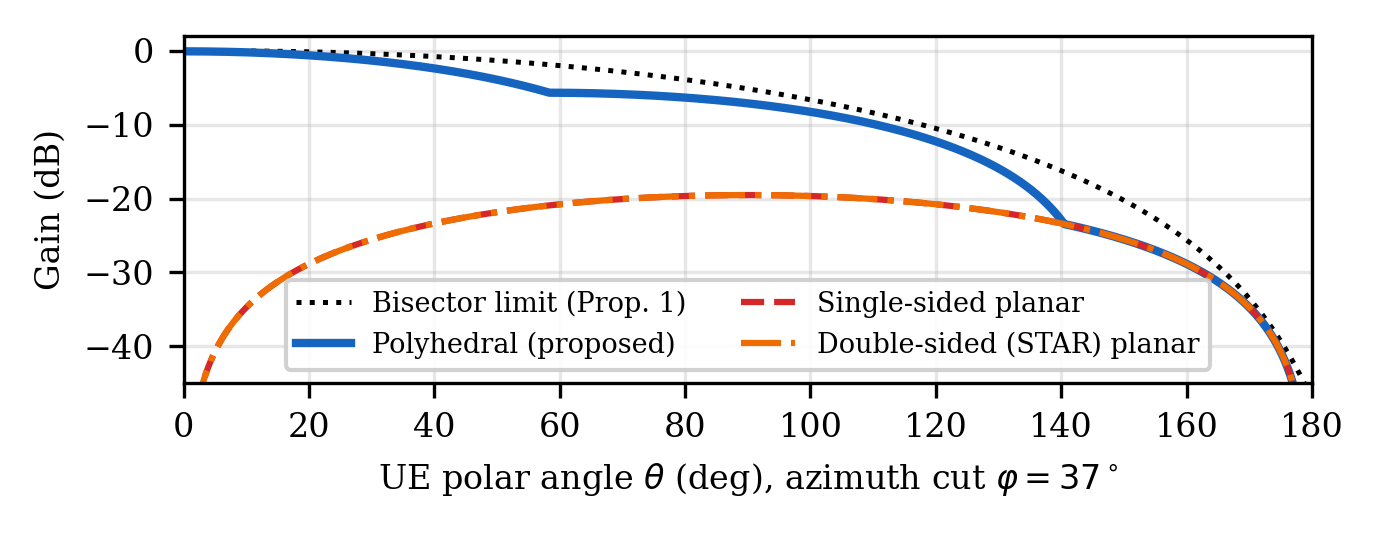}
\caption{Best facet directional gain vs.\ elevation angle.}
\label{fig:gain_elevation}
\end{figure}

\begin{figure}[t]
\centering
\includegraphics[width=3.2in, height=1.0in]{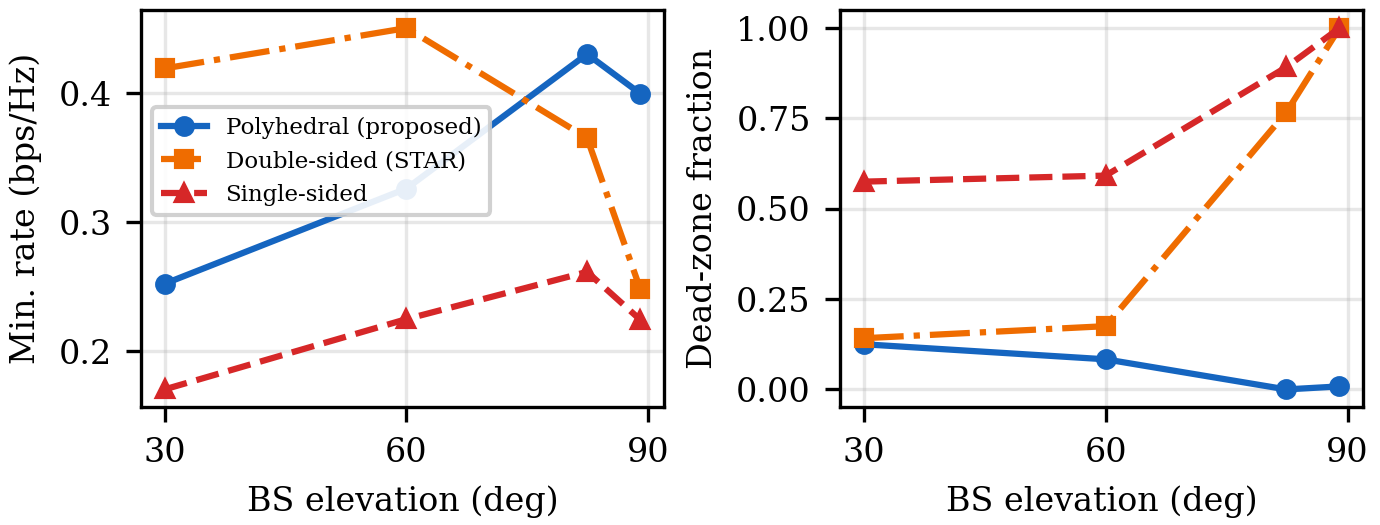}
\caption{Minimum rate and dead zone fraction versus BS elevation seen from the RIS (20 scenarios per point).}
\label{fig:elev}
\end{figure}

\begin{figure}[t!]
\centering
\includegraphics[width=3.1in, height = 0.95in]{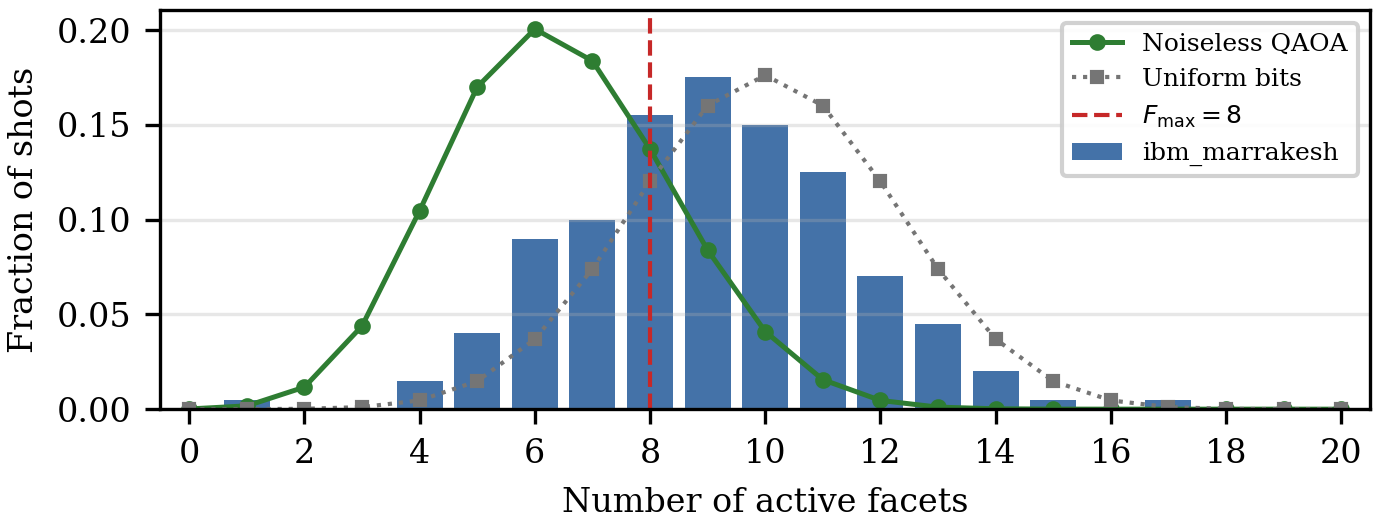}
\caption{Distribution of active facet counts on IBM quantum hardware (\texttt{ibm\_marrakesh}, corrected QUBO) against the target facet budget $F_{\max}=8$ (dashed line), noiseless QAOA, and uniform bits.}
\label{fig:hw_histogram}
\end{figure}

Facet selection formulation validation is carried out through Monte Carlo simulation and a classical simulated annealing (SA) method on the same QUBO cost surface, and compared with exhaustive search over all feasible activations.
For each Monte Carlo trial,
the UEs are randomly positioned in azimuth and an uneven distribution in elevation at a random horizontal distance away from the RIS. The path loss model used in each hop between each BS-facet and the corresponding facet-UE follows per-element free-space path loss with Rician fading.
All rates use the exact coupling \eqref{eq:exactcoupling}, $b=2$, and $p=1$.
The active facet that offers the best alignment to the UE provides service to it. The planar baselines use the same number of $F$ tiles from the vertical plane that face the BS in the standard mounting setup; the double-sided baseline uses a STAR panel during mode switching \cite{mu2022star}. The planar tiles and the last row of Table~\ref{tab:consolidated} use SA-QUBO, the other rows use greedy activation.
\begin{table}[t]
\centering
\caption{Simulation parameters.}
\label{tab:hw_execution}
\scriptsize
\setlength{\tabcolsep}{2pt}
\renewcommand{\arraystretch}{0.75}
\setlength{\aboverulesep}{0pt}
\setlength{\belowrulesep}{0pt}
\begin{tabular}{@{}l@{\hspace{2pt}}l@{\hspace{6pt}}l@{\hspace{2pt}}l@{}}
\toprule
\multicolumn{4}{c}{\textit{Simulation parameters}}\\
\cmidrule{1-4}
Facets $F{\times}N_f$ & 20${\times}$256=5120
 & UEs/scenario $K$ & 6\\
Tx / noise power & 37 / $-104$~dBm
 & Frequency & 28~GHz\\
Rician $K$ / obstruction & 5 / 60~dB
 & Path loss & Free space\\
UE distance & 20--100~m
 & BS/RIS height & 25/10~m, 2~m off\\
$\kappa_c$, $C_{\max}$ & $-32$~dB, $10^{-3}$
 & $F_{\max}$ & 8\\
SA sweeps & 400${\times}$4
 & BS antennas $M$ & 64\\
MC / sensitivity scen. & 150 / 20
 & High-elev./ISAC & 30\%\\
Near-horizon UEs & \multicolumn{3}{l@{}}{70\%, $\pm45^\circ$ from $90^\circ$}\\
\midrule
\multicolumn{4}{c}{\textit{Hardware execution (Stage 2)}}\\
\cmidrule{1-4}
Backend & \texttt{ibm\_marrakesh}
 & Qubits / layers $p$ & 24 / 1\\
Circuit depth & 963
 & Shots & 200\\
\texttt{cz} gates & 949
 & 1Q gates & 1943/1314/42\\
Variational params. & 2 (frozen)
 & Error suppression & DD, twirling\\
\bottomrule
\end{tabular}
\end{table}
Parameter values are listed in Table~\ref{tab:hw_execution}.
Table~\ref{tab:consolidated} reports the minimum UE rate across configurations. When lit at grazing incidence by the BS mentioned above, the vertical panels put 87.7\% (single sided) and 77.1\% (STAR) of the UEs into a dead zone, compared to just 0.3\% for the polyhedron. Spacing increases the minimum rate to 0.467 bps/Hz, and SA-QUBO to 0.477 bps/Hz, which is 34\% higher than the STAR panel. The STAR panel when re-tilted for the BS in question, yields 0.564 bps/Hz.
\begin{table}[t]
\centering
\caption{Ablation across 150 Monte Carlo scenarios.}
\label{tab:consolidated}
\scriptsize
\setlength{\tabcolsep}{2pt}
\renewcommand{\arraystretch}{0.85}
\begin{tabular}{lccccc}
\hline
\textbf{Configuration} & \textbf{Min.} & \textbf{Sum} & \textbf{Out.} &
\textbf{Jain} & \textbf{DZ Fr.}\\
\hline
No RIS & 0.179 & 3.42 & 0.854 & 0.677 & 1.000\\
Single-sided planar & 0.230 & 6.16 & 0.659 & 0.589 & 0.877\\
Double-sided (STAR) planar & 0.356 & 6.11 & 0.657 & 0.687 & 0.771\\
Polyhedral, 20 facets & 0.420 & 10.78 & 0.369 & 0.674 & 0.003\\
+ variable spacing & 0.467 & 10.82 & 0.360 & 0.686 & 0.003\\
+ joint activation (SA-QUBO) & 0.477 & 10.85 & 0.356 & 0.688 & 0.003\\
\hline
\end{tabular}
\end{table}
Fig.~\ref{fig:gain_elevation} shows the best facet directional gain that is computed from the facet geometry, including the incidence angle from the BS. It is compared against UE elevation angle for a fixed azimuth cut for the polyhedral, single-sided, and double-sided planar designs, together with the bisector bound of Proposition~\ref{prop:quant_loss}(i). The polyhedral surface does not exceed the bisector bound from Proposition~\ref{prop:quant_loss}(ii) by more than 3.8 dB; the vertical surfaces in full grazing illumination are 20 dB lower.
Table~\ref{tab:sensitivity} varies the Rician factor, UE number, and transmission power, contrasting the polyhedral architecture with SA-QUBO facet selection to the double-sided planar architecture. The polyhedral RIS keeps the higher minimum rate in every setting.
\begin{table}[t]
\centering
\caption{Sensitivity of minimum rate (bps/Hz): polyhedral SA-QUBO
($F_{\max}=8$) vs.\ double-sided planar RIS. $^\ast$Base condition:
$\kappa=5$~dB, $K=6$, $P_t=37$~dBm.}
\label{tab:sensitivity}
\scriptsize
\setlength{\tabcolsep}{1.8pt}
\renewcommand{\arraystretch}{0.82}
\setlength{\aboverulesep}{0pt}
\setlength{\belowrulesep}{0pt}

\begin{tabular}{@{}ccc@{\hspace{3pt}}ccc@{\hspace{3pt}}ccc@{}}
\toprule
$\kappa$ (dB) & Poly. & Planar
& $K$ (UEs) & Poly. & Planar
& $P_t$ (dBm) & Poly. & Planar\\
\midrule
0 & 0.487 & 0.395 & 2 & 3.873 & 2.965 & 31 & 0.210 & 0.158\\
3 & 0.448 & 0.372 & 4 & 1.090 & 0.591 & 37$^\ast$ & 0.422 & 0.348\\
5$^\ast$ & 0.422 & 0.348 & 6$^\ast$ & 0.422 & 0.348 & 43 & 1.023 & 0.871\\
10 & 0.371 & 0.367 & 8 & 0.297 & 0.256 & & & \\
& & & 10 & 0.198 & 0.145 & & & \\
\bottomrule
\end{tabular}
\end{table}
In Fig.~\ref{fig:elev}, the BS is placed at elevations of 30$^\circ$, 60$^\circ$, 82$^\circ$, and 89$^\circ$, when observed from the RIS. At the two lower elevations, the vertically oriented STAR panel, which is now illuminated closer to broadside, has a better minimum rate (0.419 compared to 0.252~bps/Hz at 30$^\circ$), whereas the polyhedron prevails at the two other ones. The dead zone fraction for the polyhedron does not exceed 12.5\%, but the planar panels reach 100\% (STAR).
\begin{table}[t]
\centering
\caption{Activation solvers versus the exhaustive optimum.}
\label{tab:solvers}
\scriptsize
\setlength{\tabcolsep}{2.2pt}
\renewcommand{\arraystretch}{0.82}
\begin{tabular}{lcccc}
\toprule
\textbf{Solver} & \textbf{Feasible} & \textbf{Optimal} & \textbf{Gap (dB)} & \textbf{Time}\\
\midrule
Exhaustive & 100\% & 100\% & 0 & 84~ms\\
SA-QUBO, Alg.~1 & 100\% & 95\% & 0.0009 & 167~ms\\
Greedy & 100\% & 83\% & 0.032 & $<1$~ms\\
QAOA $p=1$, noiseless$^\ast$ & 34\% & 0\% & 4.51 & --\\
Random, 200 samples & 4.8\% & 5\% & 4.22 & --\\
\texttt{ibm\_marrakesh}, $p=1^\dagger$ & 13.5\% & 0\% & 3.57 & --\\
\bottomrule
\end{tabular}
\vspace{1pt}
\parbox{\linewidth}{\centering\scriptsize $^\ast$5 instances. $^\dagger$Scenario 0, 200 shots; gap of the best feasible shot.}
\end{table}
The activation solvers are compared in Table~\ref{tab:solvers} for 150 problem instances with $K=6$. SA in Algorithm~\ref{alg1} is optimal in 95\% of these instances with the maximal gap of 0.04~dB, the greedy solver in 83\% with the maximal gap of 2.91~dB, and only 4.8\% of random activations are feasible. All 9{,}330 QUBOs solved by SA were feasible and no repairs were needed, as the Proposition~\ref{prop:qubo} guarantees. The time needed by SA to solve the instances is 0.17~s and exhaustive search needs 0.08~s, as shown in Table~\ref{tab:scale}. Dropping the coupling term affects the minimum achievable rates by 0.04\% on average and 0.2\% at most, while the first order approximation \eqref{eq:effchannel} is off from \eqref{eq:exactcoupling} by at most 0.003\%.
Table~\ref{tab:scale} provides the same comparison for chamfered $n$-gonal prisms ($F=3n+2$ facets) through an exhaustive search in streaming mode rather than storing the activations. For $n$ between 6 and 12, the number of possible activations becomes 1{,}443{,}868, and the time required by the exhaustive search is between 0.4~s and 81~s, whereas the maximum time needed by SA is only 0.43~s. For SA, 26 of 33 instances are solved optimally with the largest gap equal to 0.75~dB compared to 22 of 33 for the greedy algorithm with 1.93~dB gaps.
\begin{table}[t]
\centering
\caption{Scaling with the prism order $n$ ($F_{\max}=8$, $K=6$; times per instance).}
\label{tab:scale}
\scriptsize
\setlength{\tabcolsep}{2.2pt}
\renewcommand{\arraystretch}{0.85}
\begin{tabular}{cccrcccc}
\toprule
$n$ & $F$ & $\delta_c$ & Feasible $\boldsymbol{\alpha}$ & Exh. (s) & SA (s) & SA opt. & Greedy opt.\\
\midrule
6 & 20 & 33.7$^\circ$ & 50{,}772 & 0.4 & 0.43 & 100\% & 80\%\\
8 & 26 & 32.0$^\circ$ & 620{,}455 & 3.8 & 0.28 & 80\% & 80\%\\
10 & 32 & 31.2$^\circ$ & 1{,}443{,}868 & 16.6 & 0.31 & 70\% & 50\%\\
12 & 38 & 30.8$^\circ$ & 754{,}783 & 81.0 & 0.39 & 33\% & 33\%\\
\bottomrule
\end{tabular}
\end{table}
A two-stage QAOA validation on IBM quantum hardware was implemented.
This experiment targets the discrete step with one qubit per facet, plus four slack qubits, and exercises the whole chain, including QUBO, Ising mapping, circuit construction, pre-optimization, and transpilation.
Performing all intermediate verifications through physical hardware will consume hardware time quickly; therefore, the task is broken down into two steps, which include COBYLA constrained optimization through linear approximation using the noise-free simulator, measured by QUBO energy, which determines the value of $p = 1$ angles for the QUBO with correction in scenario 0, followed by submitting one job with fixed angles to the actual device with dynamical decoupling and gate and measurement twirling.
Table~\ref{tab:hw_execution} lists the execution details for the job.
Fig.~\ref{fig:hw_histogram} shows how many facets were active in each of the 200 shots, together with the noiseless QAOA result and uniform bits. The hardware distribution falls between noiseless QAOA result and uniform bits. 40.5\% of the samples obey the budget limit (85\% noiseless, 25\% uniform), 13.5\% are feasible (33\% and 5\%) with mean QUBO energy 86.4 (15.4 and 132.6). Best feasible sample is 3.6 dB away from the exhaustive optimal in minimum SINR. No sample achieves the optimal, thus, noise only partially keeps the bias introduced by QAOA towards feasible region. An earlier run of surrogate model formulation of the submitted problem on \texttt{ibm\_kingston} produced nearly uniform results. Noiseless QAOA increases the fraction of feasible samples from 4.8\% (random) to 34\% for $p=1$ while best of 200 samples stays still 4.5 dB away from the optimal.
\vspace{-0.6em}
\section{Conclusion}
\label{sec5}
\vspace{-0.3em}
The reformulation of the RIS to a chamfered hexagonal prism reduces the potential gain loss relative to the best reflecting facet to a closed-form expression. This is even true in the probabilistic way for the facet mounting perturbation. In case of the BS placed over the RIS, the polyhedron almost eliminates the dead zone of the vertical planar panels and achieves a minimum rate of 34\% more than the STAR panel. Furthermore, the facet selection sub-problems for the polyhedron formulation are exact QUBOs solved optimally by SA in 95\% of the time, while QAOA does not yet.
Such a solution is ideal in iterative fashion as UEs move through coherence intervals. It is evident that gains losses reduce gracefully, taking into consideration phase quantization in the hardware and mounting imperfections.
\vspace{-0.4em}
\bibliographystyle{IEEEtran}
\bibliography{cite}
\vspace{-0.3em}
\end{document}